\documentclass[a4paper,fleqn,usenatbib]{mnras}

\usepackage[T1]{fontenc}
\usepackage{ae,aecompl}

\usepackage{graphicx}	
\usepackage{amsmath}	
\usepackage{amssymb}	
\usepackage{mathrsfs}
\usepackage{bm}
\usepackage{float}
\usepackage{subfig}

\usepackage{color}
\definecolor{myColor}{rgb}{0.9,0.9,0.9}  

\title{Atmospheric Signatures of Giant Exoplanet Formation by Pebble Accretion} 

\author[Madhusudhan et al.]{
Nikku Madhusudhan$^{1}$\thanks{E-mail: nmadhu@ast.cam.ac.uk},
Bertram Bitsch$^{2}$,
Anders Johansen$^{2}$,
Linn Eriksson$^{2}$,
\\
$^{1}$Institute of Astronomy, University of Cambridge, Madingley Road, CB3 0HA \\
$^{2}$Lund Observatory, Lund University, Box 43, SE-221 00 Lund, Sweden 
}

\date{Accepted 9 May 2017. Received in original form 25 July 2016}

\pubyear{2016}

\begin{document}
\label{firstpage}
\pagerange{\pageref{firstpage}--\pageref{lastpage}}
\maketitle

\begin{abstract}
Atmospheric chemical abundances of giant planets lead to important constraints on planetary formation and migration. Studies have shown that giant planets that migrate through the protoplanetary disk can accrete substantial amounts of oxygen-rich planetesimals, leading to super-solar metallicities in the envelope and solar or sub-solar C/O ratios. Pebble accretion has been demonstrated to play an important role in core accretion and to have growth rates that are consistent with planetary migration. The high pebble accretion rates allow planetary cores to start their growth beyond 10 AU and subsequently migrate to cold ($\gtrsim$1 AU), warm ($\sim$0.1 AU- 1AU) or hot ($\lesssim$0.1 AU) orbits. In  this work we investigate how the formation of giant planets via pebble accretion influences their atmospheric chemical  compositions. We find that under the standard pebble accretion scenario, where the core is isolated from the envelope, the resulting metallicities (O/H and C/H ratios) are sub-solar, while the C/O ratios are super-solar. Planets that migrate through the disk to become hot Jupiters accrete substantial amounts of water vapour, but still acquire slightly sub-solar O/H and super-solar C/O of 0.7-0.8. The metallicity can be substantially sub-solar ($\sim0.2-0.5\times$solar) and the C/O can even approach 1.0 if the planet accretes its envelope mostly beyond the CO$_2$ ice line, i.e. cold Jupiters or hot Jupiters that form far out and migrate in by scattering. Allowing for core erosion yields significantly super-solar metallicities and solar or sub-solar C/O, which can also be achieved by other means, e.g. photo-evaporation and late-stage planetesimal accretion. 

\end{abstract} 


\section{Introduction} 
\label{sec:intro}

Atmospheric elemental abundances of solar-system giant planets have led to important constraints on their origins. For example, the observed super-solar enrichments of C, S, N, and inert gases, relative to H, in Jupiter indicate significant accretion of solid planetesimals into Jupiter's gaseous envelope \citep{owen1999,atreya2005,atreya2016}. The current Juno mission to Jupiter \citep{bolton2006,matousek2007} aims to measure the O/H abundance of Jupiter's gaseous envelope, which is crucial to constrain the formation location and accretion history of Jupiter \citep{lodders2004,mousis2012,atreya2016}. O and C are particularly important elements as they are cosmically the most abundant elements after H and He and are the major constituents of planet forming volatile molecules. The O abundance of Jupiter is currently unknown because the upper atmosphere of Jupiter (P $\lesssim$ 1 bar) has very low temperatures, T $\lesssim$ 200 K, causing H$_2$O to condense out, and be confined to the deepest layers that are inaccessible to infrared spectra ($\gtrsim$ 10 bar). Juno will measure the water abundance in Jupiter's convective outer envelope  using a microwave radiometer observing thermal emission from the deep atmosphere. Assuming the convective envelope and the atmosphere are homogeneously mixed, the global O/H ratio can be determined via the H$_2$O abundance \citep{helled2014}. 

Recent advancements in spectroscopic observations of exoplanets are making it possible to estimate precise chemical  abundances in their atmospheres, particularly for hot giant planets \citep[see e.g. recent reviews][]{madhusudhan2014a,madhusudhan2016}. The O/H, C/H, and C/O ratios are easier to measure for hot giant exoplanets than they are for solar-system giant planets \citep{madhusudhan2012}. The vast majority of extrasolar gas giants known have high equilibrium  temperatures T$_{\rm eq}$ $\sim$ 600-3000 K, thus hosting gaseous H$_2$O in their observable atmospheres accessible to spectroscopic observations. Other detectable gases include CH$_4$, CO, CO$_2$, and NH$_3$, depending on the temperature, elemental abundances, and non-equilibrium conditions \citep{lodders2002,madhusudhan2012,moses2013,tsai2016}. Measurements of such molecular abundances allow estimations of elemental abundances ratios involving H, C, O, and N. Nominal constraints on atmospheric C/H, O/H, and C/O ratios have been reported for a few exoplanets. For example, abundance estimates from dayside  thermal emission spectra of transiting hot Jupiters are indicating the possibility of a wide range of C/O ratios, ranging from carbon-rich values of C/O $\gtrsim$ 1 \citep{madhusudhan2011b,stevenson2014a} to more oxygen-rich solar-like C/O ratios of $\sim$0.5 \citep{kreidberg2014b,line2014,haynes2015}. On the other hand,  transmission spectra of a number of hot Jupiters have revealed significantly subdued H$_2$O features, implying either low metallicities and/or super-solar C/O ratios or the presence of thick clouds and hazes \citep[e.g.][]{madhusudhan2014b,benneke2015,kreidberg2015,sing2016}. Abundances have also been measured for a handful of directly-imaged planets which have thus far revealed solar-like C/O ratios \citep[e.g.][]{konopacky2013,todorov2016}. However, present constraints are only nominal given the limited spectral coverage and resolution of current instruments. More stringent constraints are expected for numerous exoplanets from upcoming observations, particular with JWST. 

The diversity of chemical abundances currently known have already motivated theoretical studies to investigate possible constraints on the formation and migration mechanisms of giant exoplanets. Motivated by early suggestions of super-solar C/O ratios in hot Jupiters \citep{madhusudhan2011b,madhusudhan2011c}, \citet{oberg2011} suggested that C/O ratios in giant planetary envelopes depend on the formation location of the planets in the disk relative to the ice lines of major C and O bearing volatile species, such as H$_2$O, CO, and CO$_2$. The C/O ratio of the gas approaches unity outside the CO$_2$ ice line as the oxygen contained in H$_2$O and CO$_2$ condenses out and only CO is in gas phase. By predominantly accreting such C-rich gas, more so than O-rich planetesimals, gas giants could, in principle, host C-rich atmospheres even when orbiting O-rich stars. In earlier studies, \citet{lodders2004} also suggested the possibility that Jupiter is C-rich and formed by accreting tar-dominated planetesimals. It may also be possible that inherent inhomogeneities in the C/O ratios of the disk itself contribute to higher C/O ratios of planets relative to the host stars \citep{kuchner2005,madhusudhan2011b,mousis2012,ali-dib2014}. The compositions of gas and solids accreted depend on the composition and thermodynamic properties of the disk at the given location which are time-dependent \citep{helling2014,marboeuf2014,thiabaud2015}. In addition, giant planets also migrate through the disk \citep{nelson2000,baruteau2014},  thereby accreting matter from different regions of the disk which then contribute to the final planetary composition. 

Recently, \citet{madhusudhan2014c} showed that the elemental abundances of hot Jupiters could potentially be used to constrain their formation and migration pathways. In their study, planetary cores that have already reached the runaway gas accretion phase migrate by Type-II migration while accreting their gaseous envelopes. The planets also accrete some planetesimals as they migrate through the disk. These planetesimals enrich the envelope of the migrating gas giant and result in super-solar carbon and oxygen abundances in the planetary envelope. Additionally, for significant planetesimal accretion, the resulting C/O ratio is sub-solar since the dominant ice composition is typically O-rich. These results imply that the sub-solar metallicities and/or super-solar C/O ratios inferred in the atmospheres of some hot Jupiters can only be explained via disk-free migration, e.g. through dynamical scattering. 

In the present study, we extend previous theoretical studies on chemical signatures of giant planet formation and migration by incorporating a more realistic planet growth and migration model and test its influence on the chemical compositions of giant planetary atmospheres. We use the planetary growth model via pebble accretion \citep{ormel2010,lambrechts2012,lambrechts2014} with subsequent gas accretion onto the planetary cores that migrate through evolving protoplanetary disks, as described in \citet{bitsch2015b}. We simulate the formation and migration of giant planets over a wide range of initial conditions, that span close-in hot Jupiters as well as Jupiter analogues at large orbital separations, i.e. cold Jupiters, and determine their chemical compositions under different assumptions for disk chemistry and for the presence of solid cores in their interiors. Based on the locations of these planets on the carbon-oxygen plane (C--O plane), we investigate the implications of using atmospheric compositions to constrain formation and migration pathways of giant exoplanets and Jupiter in the solar system. 

The paper is structured as follows. We start out by explaining our planet growth and chemical model (section~\ref{sec:models}) and show the general outcomes of our simulations in section~\ref{sec:results} for a wide range of giant planets, from hot Jupiters to cold Jupiters and Jovian analogues. We then discuss the implications of our results for chemical constraints on the formation and migration mechanisms of hot Jupiters in section~\ref{sec:hotJupiter} and of Jupiter in section~\ref{sec:Jupiter}. A discussion about our results is given in section~\ref{sec:discussion} and we summarize our results in section~\ref{sec:summary}.

\section{Models}
\label{sec:models}

\subsection{Formation Model} 
\label{sec:model_formation} 

In the core accretion scenario, a planetary core of several Earth masses forms first and then accretes a gaseous envelope to form a gas giant. Here we model the formation of the planetary core via pebble accretion, which greatly accelerates the growth speed of a planetary core \citep{johansen2010,ormel2010,lambrechts2012,lambrechts2014}, making core growth efficient also at large orbital distances. As the planet grows, it starts to open a partial gap in the disk, which hinders the pebble flux onto the planet and stops pebble accretion \citep{lambrechts2014b} and with it the accretion of solids in our model. This so-called pebble isolation mass is a strong function of the aspect ratio of the protoplanetary disk, which is given by
\begin{equation}
\label{eq:Misolation}
 M_{\rm iso} \approx 20  \left( \frac{H/r}{0.05}\right)^3 M_{\rm E} \ .
\end{equation}

As underlying disk model we use a fit to hydrodynamical simulations of a protoplanetary disk with a time-dependent mass accretion rate $\dot{M}$ and in balance between viscous and stellar heating with radiative cooling \citep{bitsch2015a}. A feature of these models is that the aspect ratio $H/r$ is smaller in the inner parts of the disk and is flared in the outer parts due to stellar irradiation, allowing the formation of larger cores in the outer disk through eq.~\ref{eq:Misolation}.

After the planet has reached its pebble isolation mass (eq.~\ref{eq:Misolation}), the accretion of solids is stopped and a gaseous envelope can contract. This contraction phase continues as long as $M_{\rm env} < M_{\rm core}$ and the corresponding contraction rates are given by 
\begin{eqnarray}
\label{eq:Mdotenv}
 \dot{M}_{\rm gas} &= 0.00175 f^{-2} \left(\frac{\kappa_{\rm env}}{1{\rm cm}^2/{\rm g}}\right)^{-1} \left( \frac{\rho_{\rm c}}{5.5 {\rm g}/{\rm cm}^3} \right)^{-1/6} \left( \frac{M_{\rm c}}{{\rm M}_{\rm E}} \right)^{11/3} \nonumber \\ 
 &\left(\frac{M_{\rm env}}{0.1{\rm M}_{\rm E}}\right)^{-1} \left( \frac{T}{81 {\rm K}} \right)^{-0.5} \frac{{\rm M}_{\rm E}}{{\rm Myr}}
,\end{eqnarray}
where $f$ is a parametric factor allowing to match the accretion rate to numerical and analytical results, which is normally set to $f=0.2$ \citep{piso2014}. The opacity in the planets envelope $\kappa_{\rm env}$ is generally very hard to determine because it depends on the grain sizes and distribution inside the planetary atmosphere. Here we use $\kappa_{\rm env} = 0.05{\rm cm}^2/{\rm g}$, which is very similar to the values used in the study by \citet{movshovitz2008} and $\rho_{\rm c}=5.5 {\rm g}/{\rm cm}^3$.

For rapid gas accretion ($M_{\rm core} < M_{\rm env}$), we follow \citet{machida2010} directly. They calculated the gas accretion rate using 3D hydrodynamical simulations with nested grids. They find two different gas accretion branches, which are given as 
\begin{equation}
 \dot{M}_{\rm gas,low} = 0.83 \Omega_{\rm K} \Sigma_{\rm g} H^2 \left( \frac{r_{\rm H}}{H} \right)^{9/2}
\end{equation}
and
\begin{equation}
 \dot{M}_{\rm gas,high} = 0.14 \Omega_{\rm K} \Sigma_{\rm g} H^2 \ ,
\end{equation}
where the effective accretion rate is given by the minimum of these two accretion rates. The low branch is for low mass planets (with $(R_{\rm H}/h < 0.3$) while the high branch is for high mass planets ($(R_{\rm H}/h > 0.3$). Additionally, we limit the maximum accretion rate to $80\%$ of the disc's accretion rate onto the star, because gas can flow through the gap, even for high mass planets \citep{lubow2006}. The growth of the planet then ends when the disc dissipates. The detailed planetary growth model is described in \citet{bitsch2015a,bitsch2015b}.

\begin{figure*}
 \centering
 \includegraphics[width=0.7\textwidth]{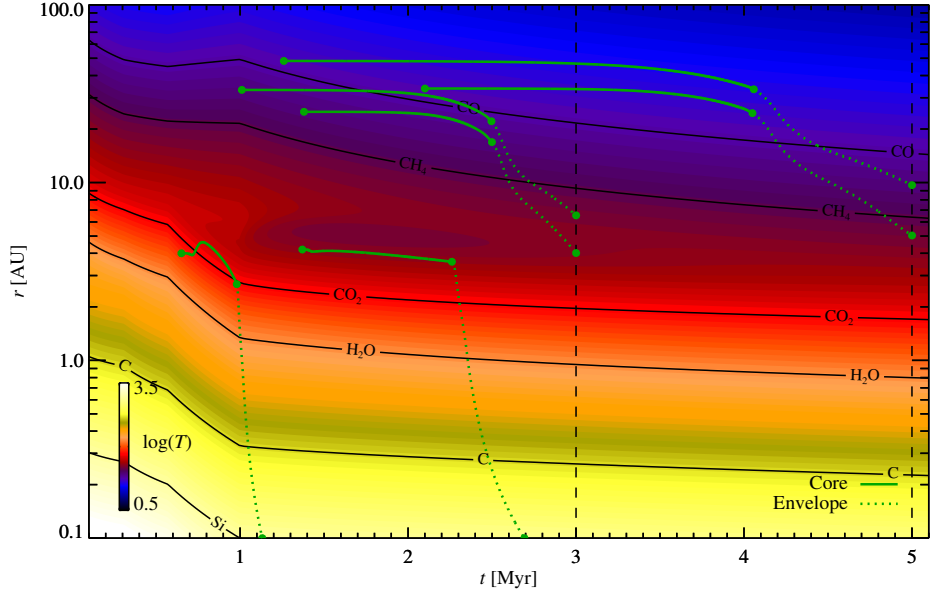}
\caption{Temperature of the protoplanetary disk as a function of time and orbital distance. The major ice lines are indicated with black solid lines. As the disk evolves, it cools and the various ice lines move closer towards the central star. Overplotted in green are orbital distances of selected planets that represent the formation pathways of cold and hot gas giants (see section~\ref{sec:model_formation})  The solid accretion phase is marked by a solid line and the gas accretion phase by a dashed line. We stop the evolution of our planets at the end of the disk's lifetime (at $3$ or $5$ Myr). Note that we stop our integration of hot gas giants when they reach the inner edge of the disk at $0.1$ AU, regardless of the total disk's lifetime.} 
\label{fig:icelines}
\end{figure*}

In Fig.~\ref{fig:icelines} we show the evolution of the temperature of the protoplanetary disk as a function of semi-major axis and time. As the disk evolves in time, it cools and thus the ice lines of the various species move towards the central star. This evolution is faster in the beginning of the disk evolution, as the inner regions of the disk are dominated by viscous heating, which diminishes as the disk evolves and loses mass in time \citep{bitsch2015a}. In the later stages of the disk's lifetime the main heating source is stellar irradiation, which only changes slightly as the star evolves in time. The viscosity of the protoplanetary disc around a solar type star is given in our model by $\alpha=0.0054$, as in \citet{bitsch2015a}. At the beginning of the disc evolution, the accretion rate is $\dot{M}=2 \times 10^{-7} M_\odot$/yr and diminishes down to $\dot{M}=2 \times 10^{-9} M_\odot$/yr after $3$ Myr and down to $\dot{M}=1 \times 10^{-9} M_\odot$/yr after $5$ Myr. We consider models with different disk life times (e.g. $3$ Myr vs. $5$ Myr). In both cases, the disc evolution is the same for the first $3$ Myr, but in the $5$ Myr case the disc is evolved for longer time before it dissipates. 

During the whole growth process, the protoplanetary disk evolves in time \citep{bitsch2015a}, and the planet migrates through the disk \citep[for a review, see][]{baruteau2014}. Low mass planets migrate through type-I migration, where the total torque $\Gamma_{\rm tot}$ is given by the sum of the Lindblad torque $\Gamma_{\rm L}$ and the corotation torque $\Gamma_{\rm C}$
\begin{equation}
 \label{eq:torque}
 \Gamma_{\rm tot} = \Gamma_{\rm L} + \Gamma_{\rm C} \ ,
\end{equation}
where the Lindblad torque is generally negative, implying inward migration. In regions of the disc, where there are strong radial gradients in entropy, the corotation torque can become positive and overpower the negative Lindblad torque, resulting in outward migration. Here we use the torque prescription by \citet{paardekooper2011}, which takes the effects of the corotation torque into account. As the time-scales for type-I migration are quite short, a fast growth is required to minimise the time spent in this phase and avoid inward migration all the way to the central star. This can be achieved by pebble accretion. However, planet migration for giant planets, in the viscous type-II migration, is still severe and they can lose a large fraction of their initial semi-major axis during their formation processes \citep{bitsch2015b}. 

We start the initial seeds at masses corresponding to the so-called transition mass where the pebble accretion rate enters the efficient Hill branch \citep{lambrechts2012}. The growth from characteristic planetesimal sizes ($\sim$100 km) to the transition mass ($\sim$0.01 M$_{\rm E}$) can happen through a combination of mutual planetesimal accretion and pebble accretion and is not modelled here, but the growth time-scale has previously been found to be on the order of 1 Myr \citep{johansen2015}. Therefore we start the transition-mass seeds 0.5-2 Myr into the evolution of the protoplanetary disk. 

In our model, we assume that the planet stops accreting solids, as soon as it reaches the pebble isolation mass (eq.~\ref{eq:Misolation}), even when it migrates through the disc. The pebbles exterior to the planet are blocked by the partial gap created by the planet \citep{lambrechts2014b}, which also exists while the planet migrates. Pebbles are subject to gas drag, which drains their angular momentum and allows a fast inward drift of the pebbles. This inward drift of the pebbles is faster than inward migration of the growing planetary cores, which thus hinders the accretion of pebbles as soon as the planet reaches pebble isolation mass. Planetesimals interior to the planetary orbit are mostly scattered away by the moving planet \citep{tanaka1999}, preventing solid accretion, but we discuss about this aspect in section~\ref{subsec:planetesimal}.

The starting planetary orbits and times were chosen in such a way that they reflect different stages of the disc evolution, but that the planets always reach 1 Jupiter mass at the end of their evolution, which limits the parameter space in starting time and initial orbital position of the planetary seed, because most of the growth tracks result in planets with different masses \citep{bitsch2015b}. Additionally, the initial semi-major axis of the planetary seeds were chosen so that they are either stranded at a few AU at the end of the disc's lifetime to resemble cold Jupiters or they migrate all the way to the inner edge of the protoplanetary disc to resemble hot Jupiters that obtained their final orbit through planet-disc interactions. What matters in the end is not the exact initial formation location in orbital distance, but the initial formation location with respect to the various ice lines. The exact orbital starting positions are given in table~\ref{tab:planetdata}.

\begin{figure}
\centering
\includegraphics[width = 0.5\textwidth]{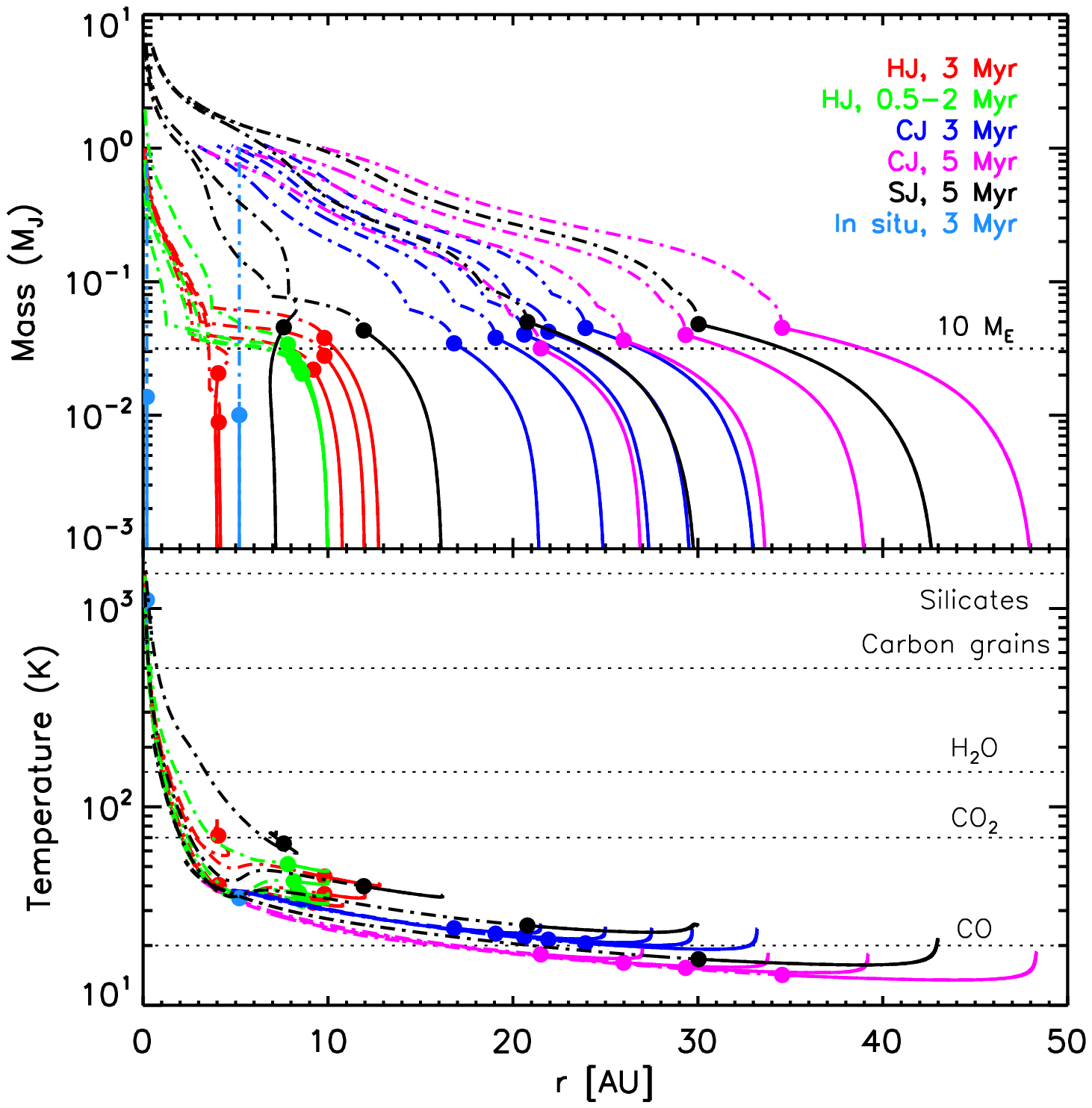}
\caption{Top: Growth tracks of planets growing via pebble and gas accretion in evolving protoplanetary disks. The x-axis shows the orbital distance in the disk. Along each track, the planet first forms a core that grows via pebble accretion (shown in thick line) until the pebble isolation mass is reached (denoted by a filled circle), following which the planet grows via gas accretion (shown in dashed lines). The different colours correspond to different final outcomes of the planet growth and migration processes, while the different lines in the same colour correspond to slightly different starting times and locations. The legend shows the different planet types and the corresponding disk life times. The model parameters for all the tracks are listed in table~\ref{tab:planetdata}. Bottom: Temperature in the disk mid-plane encountered by the planets along the different growth tracks. The condensation temperatures of the different volatile species are also shown, the intersections of which with a given growth track denote whether the species are in solid or gaseous phase. For any given growth track and a volatile molecule, the molecule is in solid (gaseous) phase for disk separations to the right (left) of the intersection.} 
\label{fig:growth_tracks}
\end{figure}

We overplot in Fig.~\ref{fig:icelines} the orbital evolution of representative cold and hot gas giants, where solid lines correspond to the core accretion phase and dashed lines to the gas accretion phase. We will refer to evolution tracks of the planetary mass (evolving by pebble accretion and gas accretion) versus the semi-major axis (evolving by migration) as {\it growth tracks}. These growth tracks are shown in Fig.~\ref{fig:growth_tracks}, where we refer to the formation location of the planet as its initial semi-major axis $r_0$ (see appendix~\ref{ap:outcomes}) corresponding to the lowest points of the growth tracks in Fig.~\ref{fig:growth_tracks}. As the planets grow and migrate, they cross several ice lines. However, since pebble accretion is very efficient, the planetary cores reach pebble isolation mass before migrating significantly \citep{bitsch2015b}. The result is that the planets mostly cross the ice lines during their gas accretion phase.

In the top of Fig.~\ref{fig:growth_tracks} the resulting growth tracks of the investigated planets are shown, while the bottom panel shows the mid-plane temperature encountered by the planets as they migrate through the disk. The thick lines in the growth tracks correspond to pebble accretion of the planetary core, while the dashed line represents the growth by gas accretion. The planets forming in the inner part of the disk have a smaller planetary core, because $H/r$ is smaller in the inner regions of the disk compared to the outer regions \citep{bitsch2015a}, which leads to a smaller pebble isolation mass and thus a smaller planetary core \citep{lambrechts2014b}. Additionally, planets that form in the later stages of the disk evolution also have smaller pebble isolation masses, as the disk becomes colder and therefore $H/r$ reduces \citep{bitsch2015a}. 

\subsection{Chemical Model} 
\label{sec:model_chem} 
Our planet growth simulations discussed in section~\ref{sec:model_formation} track the amounts of solids and gas accreted onto the forming planet, along with the local thermodynamic properties of the disk, at each time step of the planet's growth. This information is then post-processed along with the chemical prescriptions discussed below to determine the compositions of volatiles in solids and gas that are accreted into the planet and the final cumulative elemental abundances in the planet. We use here the mid-plane temperature and thus mid-plane composition of the elements in the disc and do not take vertical ice lines into account, if material were to be accreted from the top layers of the disc. However, discs heated by stellar irradiation are vertically isothermal near the mid-plane where planets grow (Bitsch et al. 2013), so that the before mentioned effect only plays a role in the inner regions of the disc.

In order to model the prominent chemical composition of the disk during the formation and migration of the planet we follow the prescription of \citet{madhusudhan2014c}. We briefly summarize the approach here. Our chemical model is based on the most dominant oxygen and carbon bearing species in the disk's mid-plane. These are H$_2$O, CO$_2$, CO, CH$_4$, silicates, and graphite grains \citep{mousis2009,oberg2011,johnson2012}. We use two different prescriptions for the chemical composition in the disk, one (case 1) based on theoretical calculations of chemistry in H$_2$-rich environments \citep{woitke2009,madhusudhan2011c} and the second (case 2) based on observations of ice and gas in protoplanetary disks \citep{draine2003,pontoppidan2006,oberg2011}. The mixing ratios (by number) of the different species as a function of the elemental number ratios (O/H, C/H, and Si/H) in the disk are shown in table~\ref{tab:chem} for both the cases. For the elemental abundances we use solar values \citep{asplund2009}: O/H $= 4.9 \times 10^{-4}$, C/H $= 2.7 \times 10^{-4}$, and hence C/O = 0.55, and Si/H $= 3.2\times 10^{-5}$. The gas in the disk mid-plane and in the planetary envelope is dominated by H$_2$ and He. The condensation temperatures (T$_{\rm cond}$) for the different species are shown in table~\ref{tab:chem}. We assume constant T$_{\rm cond}$ as the dependance of T$_{\rm cond}$ on pressure is marginal \citep{mousis2009}. 

The partitioning of the primary volatile elements, oxygen and carbon, in the solids and gas in the disk is governed by the temperature in the disk at a given time and location in the disk \citep{oberg2011,madhusudhan2014c}. Inward of the ice lines of H$_2$O, CO$_2$, and CO, the otherwise condensed forms of these species exist in their corresponding gas phases, apart from minor fractions of oxygen and carbon in refractory condensates such as silicates and graphite grains, respectively. This means that for orbital separations closer to the star than the H$_2$O ice line, C and O are predominantly in the gas phase, except for minor quantities stored in silicates and carbides. At larger separations, O is progressively depleted from the gas and moved into solid ices following the various ice lines starting with H$_2$O. Carbon on the other hand is predominantly in the gas phase until the CO$_2$ and CO ice lines. At each time step during the growth and migration of a planet in our simulation, the mass fractions of the chemical species in solids and gas accreted by the planet are computed based on the mid-plane temperature encountered by the planet and the net mass accreted in solids and gas in that time step. The mass fractions of the accreted species are subsequently converted into number fractions of the elemental abundances in the planet. For example, the mass accreted in species X (e.g. CO or H$_2$O) in gas phase in a given time step is obtained as  
 \begin{equation}
  \label{eq:Mdotgaschem}
   ~~~~M_{\rm g,X} = (2\times[X/H]\times f_{\rm H_2}\times\mu_X\times M_{\rm g, all})/\mu \ , 
 \end{equation}
where $M_{\rm g,all}$ is the total mass of gas accreted in one time-step, [X/H] is the volume mixing ratio of the species X relative to H in the gas,  $\mu_X$ is the molar mass of the species, $\mu$ is the mean molar mass of the gas, and $f_{\rm H_2}$ is the volume mixing ratio of H$_2$. For a solar abundance gas, with [He/H] = 0.085 and all H present in H$_2$, $f_{\rm H_2} = 0.85$ and $\mu = 2.3$ g/mol.

We do note that volatile chemistry in protoplanetary disks can be extremely complex and depends on a number of parameters \citep[see e.g.][]{henning2013,walsh2014,vanDishoeck2014}, especially when the abundances of complex molecules are of interest. Additionally, such disks can be significantly out of chemical equilibrium \citep{visser2011}. Nevertheless, the chemical prescriptions used in the present and previous works are chosen to account for the prominent chemistry, i.e. of the prominent molecules H$_2$O, CO$_2$, \& CO which are the dominant O and C reservoirs, while still rendering the problem tractable when including chemistry into a formation and migration model. 

{
 \begin{table*}
  \centering
 \begin{tabular}{cccc}
  \hline \hline
  Species (X) & T$_{\rm cond}$ [K]$^a$ & Case 1$^b$: X/H & Case 2$^c$: X/H \\ \hline
  CO & 20 & 0.45 $\times$ C/H (0.9 $\times$ C/H for T $<$ 70 K) & 0.65 $\times$ C/H \\
  CH$_4$ & 30 & 0.45 $\times$ C/H (0 for T $<$ 70 K) & 0 \\
  CO$_2$ & 70 & 0.1 $\times$ C/H & 0.15 $\times$ C/H \\
  H$_2$O & 150 & O/H - (3 $\times$ Si/H + CO/H + 2 $\times$ CO$_2$/H) & O/H - (3 $\times$ Si/H + CO/H + 2 $\times$ CO$_2$/H) \\
  Carbon grains & 500 & 0 & 0.2 $\times$ C/H \\
  Silicates & 1500 & Si/H & Si/H \\
  \hline
  \end{tabular}
  \caption{Condensation temperatures and volume mixing ratios of chemical species in the model disk \citep[adapted from][]{madhusudhan2014c}.
  \label{tab:chem}
  }
  {$^a$ The condensation temperatures for CO, CH$_4$, CO$_2$, H$_2$O, carbon grains, and silicates, at nominal pressures pertinent to a disk mid-plane (adopted from \citet{oberg2011} and \citet{mousis2012}). \\
  $^{b,c}$ Volume mixing ratios (i.e. by number) adopted for the species as a function of disk elemental abundances under two different prescriptions for condensate chemistry \citep[see e.g.][]{madhusudhan2014c}. Solar values are assumed for the elemental abundances O/H, C/H, and Si/H \citep{asplund2009}. The Case 2 chemistry is adopted from \citet{oberg2011} and contains carbon grains leading to more solid carbon.}
 \end{table*}
 }%

\section{Results} 
\label{sec:results} 

\subsection{Model scenarios} 
\label{sec:results_formation}

We investigate three different outcomes of giant planet formation, i) hot Jupiters, ii) cold Jupiters and iii) super-Jupiters (hot and cold), where we additionally examine the importance of the disk's lifetime. In general the initial seed locations of hot Jupiters lie interior to the initial seed location of cold Jupiters. As the disk is colder farther away from the central star, the chemical composition of the hot and cold Jupiters that migrate through planet-disk interactions will differ. More details about the different outcomes of the planet growth simulations can be found in \citet{bitsch2015b}. The exact data for all planetary simulations can be found in appendix~\ref{ap:outcomes} and Table~\ref{tab:planetdata}, and the growth tracks are shown in Fig.~\ref{fig:growth_tracks}.

The seeds of hot Jupiters start their growth early ($t_0 < 2$ Myr), for our models with total disk lifetime $t_{\rm disk} = 3$ Myr and start in the inner part of the protoplanetary disk ($r_0 < 15$ AU). They migrate all the way to the inner edge of the disk ($r_{\rm f} = 0.1$ AU) during the lifetime of the disk (red and green lines in Fig.~\ref{fig:growth_tracks}). Note that the growing planets indicated by the green lines in Fig.~\ref{fig:growth_tracks} are all starting at the same location, but have different starting times. The main phase of gas accretion happens in the very inner regions of the disk ($r\lesssim3$ AU), where viscous heating  ($\alpha=0.0054$) dominates and the disk is thus very hot. This means that these planets will accrete most of the volatiles (CO, CO$_2$, and H$_2$O) in gas phase. 

The seeds of cold Jupiters in a disk with lifetime $t_{\rm disk} = 3$ Myr also form early ($t_0 < 1.5$ Myr), but in the outer regions of the protoplanetary disk ($r_0 \gtrsim 15$ AU) as indicated by the blue lines in Fig.~\ref{fig:growth_tracks}. They need to form farther away from the central star in order to avoid inward migration all the way to the central star. The outer regions of the disk are cooler and thus several oxygen-rich volatile species, e.g. H$_2$O and CO$_2$, are in solid form. This means that during the gas accretion phase these volatiles are not accreted and the planets therefore only accrete CO gas, resulting in a super-solar C/O ratio of 1.

The seeds of cold Jupiters that form in a disk with longer lifetime ($t_{\rm disk} = 5$ Myr) have to start at a  slightly later time ($1$ Myr $< t_0< 2.5$ Myr) and farther away from the central star ($r_0 > 25$ AU) than the cold Jupiter counterparts that formed in the disk with shorter lifetime. They are marked in magenta colour in Fig.~\ref{fig:growth_tracks}. As the disk cools down during its lifetime (see Fig.~\ref{fig:icelines}), we expect that those planets will accrete even less volatiles in their envelope during their formation compared to their counterparts in the 3 Myr disk, as also CO can be frozen out into solids and can thus not be accreted in gaseous form. Those planets can therefore have significantly sub-solar oxygen and carbon abundances, in addition to having a C/O ratio of 1. 

The super-Jupiter planets in our simulations are planets of several Jupiter masses (indicated by the black colour in Fig.~\ref{fig:growth_tracks}). We further distinguish between hot and cold super-Jupiter planets. As super-Jupiter planets need to grow for a longer time, we only investigate their growth process in disks with lifetimes of $5$ Myr. The formation pathway of the super-Jupiters is similar to the formation pathways of the regular Jupiter mass planets. However, these planets migrate more and are thus accreting more gas from the inner regions of the protoplanetary disk. We therefore expect their atmospheres  to be more enriched in oxygen and carbon than their Jupiter mass counterparts.

Additionally we also investigate the formation of hot and cold Jupiters that form in-situ, by just accreting local material without migrating through the disk. The composition of the accreted material can therefore only change if the disk cools significantly and an ice line moves across the planetary orbit. The chemical composition of these planets is therefore directly linked to the initial formation location of the planetary seeds and to the disk's temperature at that location.

\subsection{Chemical Signatures} 
\label{sec:results_chem} 

Here we discuss the chemical abundances in the atmospheres of the three classes of giant planets discussed in section~\ref{sec:results_formation}. The envelope composition of a gas giant planet is governed by the contributions of both gas accretion as well as the accretion of solids in the envelope during formation. In the standard pebble accretion mechanism, as  described above, in the initial stages of formation the solids are accreted as pebbles which lead to core formation. Once the core reaches a critical mass, namely the pebble isolation mass, pebble accretion ceases following which only gas is accreted to form the giant planet. This mode of giant planet formation leads to two very different pathways for the effect of solids on the envelope composition. In the standard scenario, assuming the core is intact and isolated from the rest of the planet, the envelope composition is entirely governed by the composition of the accreted gas alone. On the other hand, if the core is eroded over time the envelope could be enriched post-formation with heavy elements from the core mixing into the envelope.

\begin{figure}
\centering
\includegraphics[width = 0.5\textwidth]{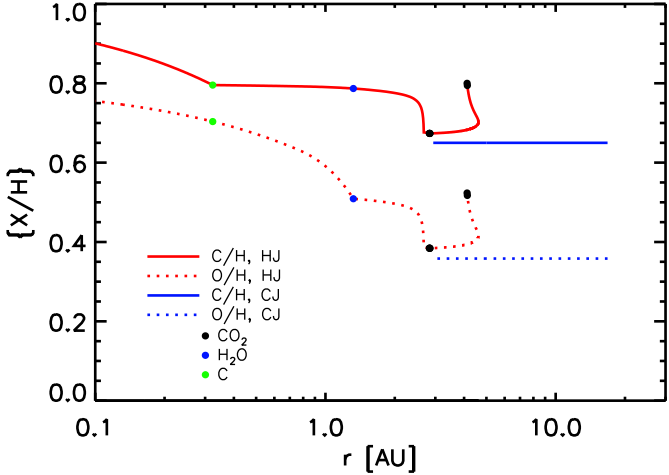}
\caption{Evolution of O and C abundances for two representative examples. The y-axis shows the  elemental abundance ratio X/H relative to solar value, where X is either O (dashed line) or C (solid line), of the planetary envelope as a function of semi-major axis for an example of a hot Jupiter (red) and cold Jupiter (blue). The red curves correspond to the innermost red HJ growth track in \ref{fig:growth_tracks} starting at 4 AU and the blue curves correspond to the innermost blue CJ growth track in \ref{fig:growth_tracks} starting at 21.5 AU. As the planets move through the disk and cross different ice lines (indicated by coloured dots), the composition of the accreted gas changes and with it the \{C/H\} and \{O/H\} ratios. Strong changes of the \{C/H\} and \{O/H\} ratios happen after the planet crosses ice lines. These  examples have been calculated assuming case 2 chemistry for the condensation as shown in Table~\ref{tab:chem}.}
\label{fig:ratio}
\end{figure}

In Fig.~\ref{fig:ratio} we show the \{C/H\} and \{O/H\} ratios of two representative examples, one hot Jupiter and one cold Jupiter, as they grow and migrate through the disk. These two examples correspond to the innermost growth tracks in each category shown in \ref{fig:growth_tracks}: the red curves correspond to the innermost red growth track in \ref{fig:growth_tracks} starting at 4 AU and the blue curves correspond to the innermost blue growth track in \ref{fig:growth_tracks} starting at 21.5 AU. The seed of the cold Jupiter forms at $r>r_{\rm CO_2}$, but as it migrates it never crosses the CO$_2$ ice line and thus only accretes CO gas besides hydrogen. As we use the chemical model 2 for the tracks shown in Fig.~\ref{fig:ratio}, the \{C/H\} ratio is always 0.65 and \{O/H\} is $\sim 0.35$, leading to a C/O of 1.86$\times$ solar, i.e. 1.0. 

The seed of the hot Jupiter, on the other hand forms at $r<r_{\rm CO_2}$, but crosses the CO$_2$ ice line as it grows and is caught in the region of outward migration\footnote{Outward migration occurs in disks where the local radial gradients of entropy are such that the entropy driven corotation torque can overpower the inward directed Lindblad torque. In our disk models this is the case in regions where H/r decreases radially close to the water ice line, see  \citet{bitsch2015a}.}. As the disk evolves, the region of outward migration moves closer to the central star, allowing the planet to cross the CO$_2$ ice line again (when the black dot in Fig.~\ref{fig:ratio} is crossed again) as the planet starts to contract its envelope. The planet then undergoes runaway gas accretion and is not contained in the region of outward migration any more. It then migrates inwards and crosses the H$_2$O ice line when it is roughly $\sim$ 20\% of Jupiter's mass and the carbon grains ice line closer in to the star. At this point, the planet's oxygen and carbon abundances increase dramatically as it migrates all the way to the central star. At it's final location at the disk's inner edge, its \{C/H\} is roughly solar (0.9$\times$), but its \{O/H\} ratio is slightly more depleted compared to solar (0.76$\times$), leading to a C/O of 0.66. Note here that when the planet crosses the water ice line, it will still not accrete a solar ratio of \{O/H\}, because part of the oxygen is bound in silicates (see table~\ref{tab:chem}).

\begin{figure*}
\centering
\includegraphics[width = 0.75\textwidth]{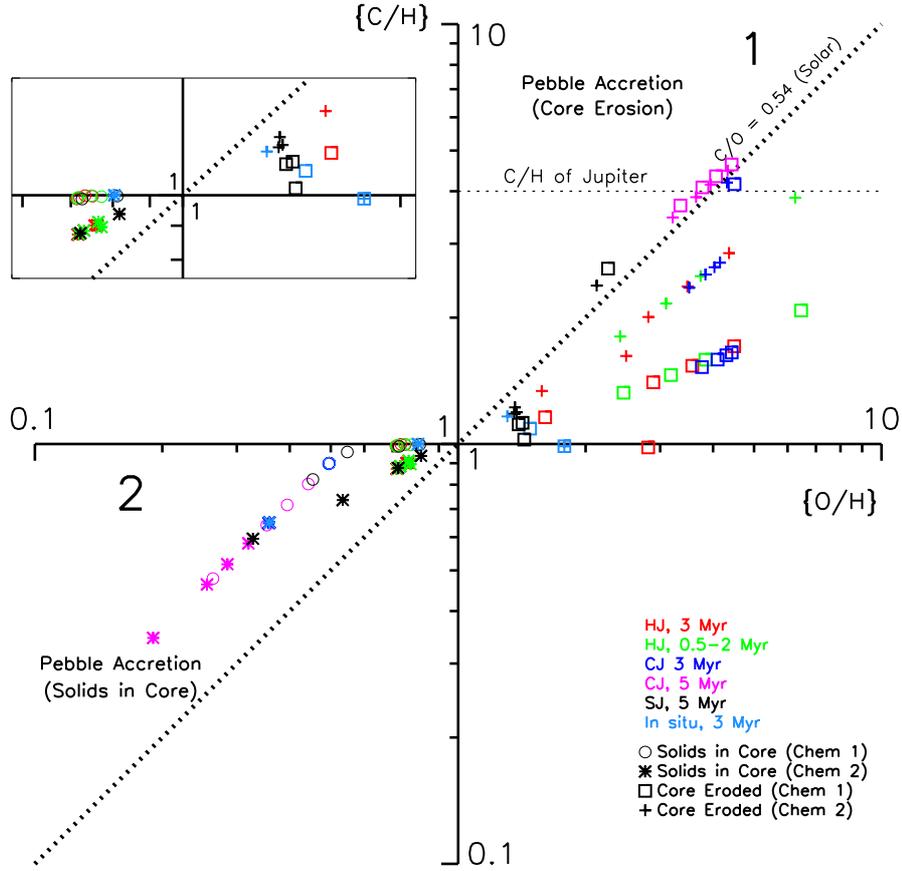}
\caption{Oxygen and carbon abundances of our models of giant planets formed by pebble accretion. The axes show O/H and C/H ratios in the planets relative to solar values assumed for the disk. Each coloured symbol corresponds to a model track of the same colour shown in Fig~\ref{fig:growth_tracks}. The circles show models in the standard pebble accretion paradigm assuming that the solids remain in the core and do not affect the composition of the envelope. Under this condition all the models from Fig.~\ref{fig:growth_tracks} result in solar or sub-solar C/H and O/H ratios, thereby occupying the bottom-left quadrant of the C--O plane as shown in the Figure. The circles and asterisk symbols in this quadrant show the chemical abundances for the same growth tracks but with two different assumptions for the disk chemistry listed in table~\ref{tab:chem}, i.e. case 1 and case 2, respectively. The squares and plus symbols in the top-right quadrant shows the same models as the bottom-left quadrant but with the assumption that all the solids that are accreted into the core eventually mix into the envelope due to core erosion. The inset in the top left shows an extended view of the central region of main plot. The values corresponding to the models are shown in Tables~\ref{tab:planetchem} and \ref{tab:planetchem_oberg}.}
\label{fig:COplane}
\end{figure*}

\subsubsection{Standard Pebble Accretion - Sub-solar Metallicities and Super-solar C/O ratios} 

The main consequence of giant planet formation by pebble accretion is the low metallicities in the envelope, as shown in the lower-left quadrant of the C--O plane in Fig.~\ref{fig:COplane}. The key assumption here is that the solids that initially accreted via pebbles to form the core stay within the core and do not mix in the envelope. In this scenario, the metallicity of the envelope, and hence the atmosphere, is determined only by the composition of the accreted gas. For a solar composition disk, the maximum metallicity possible for any element (e.g. C, O, etc.) in the gas in any part of the disk is solar metallicity. Therefore, the elemental abundances of the gaseous envelope accreted by the planet are solar or sub-solar, depending on the location in the disk from where the gas was accreted. 

Here, we discuss the O/H, C/H, and C/O ratios of planets formed via pebble accretion for the different model scenarios and their growth tracks  discussed in section~\ref{sec:results_formation}. For each model we compute the chemical compositions for two different prescriptions of the ice chemistry (case 1 and case 2,  listed in Table~\ref{tab:chem}). All the compositions are shown in Fig.~\ref{fig:COplane}; results from case 2 chemistry from \citet{oberg2011}, which is more carbon-rich, are identified by the asterisk and plus signs. The values for case 1 and case 2 chemistry are also shown in Tables~\ref{tab:planetchem} and \ref{tab:planetchem_oberg}, respectively. Here, we discuss results from case 1 chemistry for simplicity. As seen from Fig.~\ref{fig:COplane}, the results from case 2 chemistry closely follow the trends from case 1 chemistry but with a generally stronger depletion or enhancement of carbon in the lower-left and upper-right quadrants, respectively. 

For each element, the gas metallicity in the disk becomes sub-solar beyond the ice lines of molecules containing the corresponding species \citep[e.g.][]{madhusudhan2014a}. Given their high condensation temperatures, silicates (e.g. MgSiO$_3$) remain in solid phase for much of the disk except the innermost regions ($\lesssim$0.1 AU). Consequently, the disk gas beyond $\sim$0.1 AU is depleted in O by $\sim$20\% even within the H$_2$O ice line and is further depleted at each subsequent volatile ice line outward in the disk. Therefore, as shown in Fig.~\ref{fig:COplane}, planets accreting most of their envelope within the H$_2$O ice line, e.g., hot Jupiters, have solar C/H ratio (i.e. \{C/H\} = 1) and slightly sub-solar O/H ratio (\{O/H\} $\sim$0.8). The corresponding C/O ratio is $\sim$1.25$\times$Solar C/O, i.e. C/O $\sim$0.7. 

For planets forming beyond the H$_2$O ice line, with increasing distance in the disk the atmospheric metallicity becomes progressively sub-solar and the C/O ratio becomes progressively super-solar, as also found in previous studies \citep{oberg2011,madhusudhan2014c}. As shown in Fig.~\ref{fig:COplane}, these planets have sub-solar O/H and C/H ratios. The sub-solar abundances are a consequence of the volatile molecules that are trapped in solid ices beyond their respective ice lines thereby depleting the gas of the corresponding elements. The condensation temperature of H$_2$O is highest, followed by CO$_2$ and then CO, so the gas is more depleted in oxygen relative to carbon for planets forming between the H$_2$O and CO ice lines. 

The final chemical composition of a planet is a cumulative effect of the gas accreted from different regions of the disk along the planet's migration path and, hence, may not be representative of any one particular zone. Overall, we find that giant planets forming beyond the H$_2$O ice line can reach metallicities as low as $\sim$0.2$\times$ solar and C/O ratios close to unity, for initial seed formation locations between $\sim$10 and 50 AU. Note that C/O ratios of unity can be achieved even if the planets originate outside the CO ice line, because even though the gas has almost no C or O based volatiles outside the CO ice line the planet is migrating inward and accretes most of the gas within the CO ice line. Thus, the accreted gas will have a C/O ratio $\sim$1 if the gas is accreted mostly between the CO$_2$ and CO ice lines. 
 
Our models suggest that pebble accretion and disk migration result in generally sub-solar C/H and O/H ratios and super-solar C/O ratios that are a strong function of their initial formation location and migration path in the disk. The hot Jupiters in our models almost all originate within $\sim$20 AU and accreted a large fraction of their mass within the H$_2$O ice line, where most of C and O are in gas phase and silicates are the only condensates. Thus our hot Jupiter models have nearly solar C/H ratios, slightly sub-solar ($\sim$0.7-0.8$\times$solar) O/H ratios, and only slightly super-solar C/O ratios ($\sim$0.7), assuming case 1 chemistry for the disk. On the other hand, the cold Jupiters in our models can form much farther out and lead to substantially lower O/H and C/H ratios and high C/O ratios ($\sim$1) as discussed above. We note again that in this standard scenario we assume the pebbles are isolated in the core and do not contribute to the envelope metallicity. In reality, it may be possible that some pebbles still accrete after the isolation mass is reached and/or are ablated in the proto-atmosphere during accretion \citep{venturini2016} and enhance the metallicity of the accreted envelope. 

Our results suggest that a multitude of formation-migration pathways may be responsible for explaining the chemical diversity in giant exoplanets. The chemical signatures discussed above were based on the assumption of disk migration alone, while disk-free migration after formation could transport our cold Jupiters to close-in orbits. These scattered cold Jupiters would now be observable as hot Jupiters and could explain the sub-solar O/H ratios and super-solar C/O ratios suggested for some hot Jupiters \citep[e.g.][]{madhusudhan2011,madhusudhan2014a}. Thus, our models with standard pebble accretion predict solar or sub-solar metallicities for hot Jupiters irrespective of migration mechanism and super-solar C/O ratios ranging between $\sim$0.7-1.0 depending on whether the planet migrated through the disk or disk-free. On the other hand, the observations of super-solar metallicities and solar C/O ratios in giant exoplanets \citep[e.g.][]{kreidberg2015} would need alternate explanations, e.g., late-stage planetesimal accretion, as discussed in section~\ref{sec:hotJupiter}. Below, we discuss the implications of the same models of formation by pebble accretion discussed above but allowing the pebbles in the core to gradually erode and mix into the envelope. 

\subsubsection{Pebble Accretion with Core Erosion -  Super-solar Metallicities and Sub-solar C/O ratios} 
\label{subsec:erosion}

Solid cores of giant planets can undergo significant erosion and the material redistributed to the envelope \citep{wilson2012}. In order to test the effect of this process, here we investigate the chemical consequences of 100\% core erosion on the atmospheric composition of giant planets in our models,  where all chemical elements that were originally trapped in the core are evenly distributed into the planetary atmosphere. The resulting O/H and C/H ratios are shown in the top-right quadrant of Fig.~\ref{fig:COplane}. We find that assuming 100\% core erosion in our models leads to super-solar O/H and C/H ratios up to 4$\times$solar and C/O ratios that are nearly solar ($\sim$0.5) or sub-solar. The origin of the super-solar metallicity is due to excess accretion of  solids, the compositions of which depend on the initial formation location of the seed of the planet. Contrary to gas accretion, which happens as the planet migrates across the disk, the solid accretion and core growth happens almost locally near the initial formation location of the core (see Fig.~\ref{fig:icelines} and Fig.~\ref{fig:growth_tracks}). 

The final composition of the planet is given by the total contribution from both the accreted gas and solids. The solid composition is oxygen-rich between the H$_2$O and CO$_2$ ice lines, where only silicates and H$_2$O ice are available, and becomes increasingly carbon-rich father out in the disk as CO$_2$ and CO ices become available. A maximum C/O ratio of solar value (0.55) is reached in solids when all O and C bearing volatiles are in solids beyond the CO ice line. Therefore, while the enhancement of the O/H ratio generally depends on the amount of solids accreted, the C/H and C/O ratios depend on the location in the disk from where the solids are accreted. For example, the top-right quadrant of Fig.~\ref{fig:COplane} shows two broad regions of elemental abundances for each chemical case, all of which span a similar range in O/H ratio but differ in the C/H ratio. The magenta colored models (cold Jupiters in disks with $t_{\rm disk}=5$ Myr) with C/O$\sim$0.5 indicate planets that formed farthest in the disk, beyond the CO ice line, and hence accreted solids of nearly solar C/O ratio. On the other hand, planets forming between the H$_2$O and CO ice lines (e.g., the red and green models indicating hot Jupiters) accrete predominantly H$_2$O and some CO$_2$ in solids leading to high O/H ratios but low C/H ratios and, hence, low C/O ratios (e.g. as low as 0.2).  

The C/H and O/H ratio in the envelopes of super-Jupiter planets is not increased as much by core erosion compared to the normal Jupiters. The reason is simply that oxygen and carbon of a similar sized core is eroded in a much larger planetary atmosphere, yielding a smaller increase in the \{C/H\} and \{O/H\} ratios.

\section{Implications for the formation and migration of hot Jupiters}
\label{sec:hotJupiter}

Our results have important implications for constraining the formation and migration mechanisms of hot Jupiters based on their observed atmospheric O/H, C/H, and C/O ratios. The elemental abundances resulting from our models, as shown in Fig.~\ref{fig:COplane} and Tables~\ref{tab:planetchem} \& \ref{tab:planetchem_oberg}, span four distinct formation-migration pathways of hot Jupiters: (1) Formation of a hot Jupiter by standard pebble accretion, with no core erosion, and migration through the disk to present orbit at $\sim$0.1 AU separations, (2) Formation of a cold Jupiter by standard pebble accretion far out in the disk and migrating to close-in orbits later by dynamical scattering (i.e. disk-free migration), (3) Formation of a hot Jupiter by pebble accretion and disk migration but with the core eroded over time, and (4) in-situ formation of a hot Jupiter. In what follows, we discuss the observable implications of each scenario. 

Firstly, our model hot Jupiters formed via standard pebble accretion (with intact cores and migrating through the disk) are typically characterized by solar or sub-solar metallicities and super-solar C/O ratios. These systems are shown in red and green in quadrant 2 in Fig.~\ref{fig:COplane}. These systems have nearly solar C/H ratios and marginally sub-solar O/H ratios ($\sim$0.7-0.8 $\times$solar), leading to slightly super-solar C/O ratios of $\sim$0.7-0.8; the solar C/O ratio is $\sim$0.55. We find this trend irrespective of the assumptions of the condensate chemistry from our two cases (Table~\ref{tab:chem}). These results follow because the hot Jupiters that migrated through the disk to their present orbits needed to have formed within $\sim$15 AU separations, and accreted most of their mass in gas within the H$_2$O ice line. Therefore, the only condensates condensing out of the gas are silicates, thereby leading to only slightly sub-solar O abundance, as discussed in section~\ref{sec:results_chem}. 

Secondly, it is possible that present-day hot Jupiters were initially formed at large orbital separations as cold Jupiters and then migrated to close-in orbits via dynamical encounters (i.e. disk-free migration). In that case, the chemical signatures we see in the hot Jupiters would be those of our cold Jupiter models, since the chemistry is expected to be unchanged in disk-free migration. In this scenario, and continuing the assumption of intact cores, we find that the resulting planets can have significantly sub-solar metallicities and super-solar C/O ratios: O/H of $\sim$0.2-0.5$\times$ solar, C/H of $\sim$0.5-0.9$\times$ solar, and C/O ratios close to unity. This is because most of the gas mass in these cases is accreted between the CO$_2$ and CO ice lines due to which only CO is in gas phase. 

Thirdly, if we assume that the cores of the planets discussed above are eroded over time, then we predict very different chemical abundances for hot Jupiters. In this scenario, as discussed in section~\ref{sec:results_chem}, we find a wide range of O and C abundances but following a general trend of super-solar metallicities and solar or sub-solar C/O ratios. The specific abundances in the models, depending on the specific formation tracks and assumptions for condensate chemistry, span values of O/H $\sim$1-5 $\times$ solar, C/H $\sim$1-4$\times$solar, and C/O $\sim$ 0.2-0.6. These metal enhancements follow the general fact that addition of the solids into the envelope enhances the metallicity of the envelope, and since the solids are mostly oxygen-rich they tend to decrease the C/O ratio of the envelope. Moreover, the further in the disk the planetary core forms the more the enhancement in O and C because more of the volatiles are in solid phase. We also note that the O/H enhancement is more pronounced for lower planet masses as the relative contribution of the solids to the envelope is higher. For example, the green points with the highest O/H in quadrant 1 of Fig.~\ref{fig:COplane} correspond to the lowest planet mass. 

Finally, in-situ formation of hot Jupiters has recently  experienced additional attention in the literature \citep{batygin2016}, where the basic idea is that the planetary seed forms close to the central star and subsequently accretes gas from the surrounding disk (light blue in Fig.~\ref{fig:growth_tracks}). The accreted solid material in the hot inner regions of the disk will result in a core that consists mostly of silicates and only to a tiny fraction of oxygen, as most of the oxygen and all of the carbon harbouring species are in gaseous form. This also applies if the planetary core is formed via the accretion of pebbles that drift through the disk \citep{chatterjee2015}, as e.g. H$_2$O ices evaporate before reaching the very inner regions of the disk. On the other hand, the accreted gas contains nearly all oxygen and all carbon harboring species, resulting in a C/O ratios close to solar, see Fig.~\ref{fig:COplane} and Tables~\ref{tab:planetchem} \& \ref{tab:planetchem_oberg}. Our model of in-situ formation of a hot Jupiter results in a solar C/H and an O/H that is slightly sub-solar (0.8$\times$solar, for no core erosion) or slightly super-solar (1.5$\times$solar, for core erosion) and, accordingly, a C/O ratio slightly super-solar (0.7) or sub-solar (0.4). 

Based on the above findings the diversity of observed elemental abundances in hot Jupiters might indicate a diversity of their formation-migration pathways. Hot Jupiters that are found to have only slightly sub-solar abundances (0.7-1 $\times$ solar) and slightly super-solar C/O ratios (0.7-0.8) would be suggestive of formation by the standard pebble accretion mechanism with intact cores and migration through the disk or in situ formation. On the other hand, hot Jupiters with significantly low metallicities ($\lesssim$ 0.7$\times$ solar) and super-solar C/O ratios ($\sim$1) would be strongly indicative of disk-free migration, consistent with predictions from \citet{madhusudhan2014c}. On the contrary, hot Jupiters with super-solar metallicities and solar or sub-solar C/O ratios  could be the result of a multitude of processes, including core erosion, disk migration, and possibly late stage significant planetesimal accretion and/or photoevaporation as discussed in section~\ref{sec:Jupiter}. Late stage planetesimals impacting a Jovian-like planet are expected to undergo substantial ablation in the planetary atmosphere thereby enriching its volatile abundances \citep[e.g.,][]{mordasini2015,pinhas2016}. Current observations of hot Jupiters are already revealing a diversity of compositions ranging from sub-solar to super-solar O/H abundances \citep[e.g.][]{madhusudhan2014a,kreidberg2015} and a wide range of C/O ratios \citep[e.g.][]{madhusudhan2011b,line2014,benneke2015}, suggesting a diversity of possible origins of hot Jupiters. 

\section{Implications for the formation of Jupiter} 
\label{sec:Jupiter}

Jupiter's atmosphere is enriched in carbon compared to solar value by a factor of $\sim$ 4 $\pm$ 1 \citep{atreya2016}, which we can achieve in our model by eroding the planetary core. Another possibility of increasing the oxygen and carbon in the planetary envelope is through the additional accretion of solids.  As the accretion of pebbles is hindered by the planet's feedback on the disk \citep{lambrechts2014b}, only the accretion of planetesimals that then get ablated in the planetary envelope \citep[e.g.,][]{mordasini2015,pinhas2016} can increase the \{C/H\} and \{O/H\} ratios. Yet another possibility is the accretion of gas that is already enriched in heavy elements as the gas envelope is accreted. We will discuss both possibilities in the following.

\subsection{Early planetesimal accretion}
\label{subsec:planetesimal}

In our simulations, Jupiter forms in the outer disk and migrates towards its final position at $\sim 5$ AU (blue and magenta lines in Fig.~\ref{fig:growth_tracks}). Most of its migration happens when the planet is in the gas accretion phase. As the planet migrates, it would encounter planetesimals that formed interior to its birth orbit, an effect which we do not model in our simulations. If core erosion is not taken into account, then the mass of planetesimals that needs to be accreted into the envelope is similar to the mass of the core\footnote{100\% core erosion lifts our simulated Jupiters to the \{C/H\} that is observed, Fig.~\ref{fig:COplane}}. In our simulations this corresponds to 10-15 Earth masses (Fig.~\ref{fig:growth_tracks}).

N-body simulations of migrating planets have shown that only a few \% of the planetesimals are actually accreted by a migrating planetary core \citep{tanaka1999}. If we were to invoke that {\it all} the pebbles in the disk at the beginning of the growth of the planet have turned into planetesimals, then inside the initial planetary orbit and it's final location about $30$ Earth masses of planetesimals exist in our model. The efficiency of planetesimal accretion during migration therefore would have to be larger than 30\%, which is much larger than predicted by the N-body simulations of \citet{tanaka1999}. 

An alternative formation scenario of Jupiter is given by the Grand Tack scenario \citep{walsh2011}. There, Jupiter migrates into the inner solar system (down to 1.5-2.0 AU) and then outwards again with the help of Saturn. During this time, Jupiter crosses the asteroid belt, which is thought to be more massive in the early solar system evolution, twice. However, during this migration path, Jupiter just accretes about $\sim 0.5$ Earth masses in solids (S. Jacobson, private communication). We therefore conclude that early accretion of planetesimals during Jupiter's formation is likely not responsible for the super-solar abundance in carbon in Jupiter's atmosphere.

\subsection{Late stage planetesimal accretion}
In the Nice model \citep{tsiganis2005}, Jupiter and Saturn start out in resonance, which is disturbed by an outer belt of planetesimals\footnote{This planetesimal belt normally starts at around $30$ AU, which is outside of the starting position of about half our growth tracks that produce Jupiter type planets (see Fig.~\ref{fig:growth_tracks}).}. The planetary configuration becomes unstable and Jupiter and Saturn reach their final orbital position. This instability generates a bombardment of the inner terrestrial planets know as the Late Heavy Bombardment. During this bombardment Jupiter also accretes some planetesimals into its atmosphere. Planetesimals accreting into a gas giant planet ablate rather efficiently in the planetary envelope, particularly the volatile ices, thereby enhancing its volatile abundances \citep[][]{mordasini2015,pinhas2016}. However, the Late Heavy Bombardment only amounts to $\sim 0.15$ Earth masses \citep{matter2009}, which is clearly not enough to explain the enrichment in Jupiter's atmosphere. Another possibility is one of a substantial planetesimal accretion history post formation beyond the Late Heavy Bombardment. 

\subsection{Disc photoevaporation}
Another mechanism to increase the carbon and oxygen abundance in Jupiter's atmosphere can be realised when the accreted gas is already enriched in carbon and oxygen. This can be achieved in the late stages of the disk evolution, when photoevaporation starts to become important \citep{guillot2006}. Volatile gases condense onto existing grains in the outer disk, which allows them to settle to mid plane. Photoevaporation first erodes the upper layers of the disk which consists mainly of hydrogen. As then the gas and grains move inwards, eventually the evaporation temperature of the volatile gases that condensed onto the grains is reached and the volatiles are released into the disk. This effect increases the volatile abundance to super-solar levels. If this gas is now accreted onto Jupiter, it can naturally explain the enrichment of all volatiles (e.g. C and O), but also the super solar abundance of noble gases in the atmosphere of Jupiter.

In our model, Jupiter's core forms in the very cold parts of the disk, where $T<30$K (Fig.~\ref{fig:growth_tracks}). At these temperatures the noble gases are condensed out onto pebbles \citep{guillot2006}, which are the building blocks of the core and are thus directly incorporated into the planetary core. Eroding the core can thus not only explain the amount of carbon in Jupiter's atmosphere (Fig.~\ref{fig:COplane}), but can potentially also explain the enrichment in noble gases.

However, if the core of Jupiter were to form in-situ at 5.2 AU (light blue in Fig.~\ref{fig:growth_tracks} and Fig.~\ref{fig:COplane}), disk photoevaporation is likely the only way of explaining the large carbon abundance. The core forms at $r>r_{\rm CO_2}$ and $r<r_{\rm CO}$, which indicates that the core does not consist of CO ice, which is the most abundant carbon bearing species in our chemical models (table~\ref{tab:chem}), making the core less abundant in carbon compared to the Jupiter analogues that form far out in the disk. Therefore, even when taking complete core erosion into account, the C/H ratio of Jupiter in our solar system cannot be matched considering  formation by pebble accretion alone. And, since the accretion of planetesimals also cannot solely account for the carbon abundance as discussed above, photoevaporation remains the only explanation for Jupiter's super solar carbon abundance if Jupiter was formed  in-situ. 

\subsection{Predictions for Jupiter's composition}

Jupiter in our own solar system is enriched in carbon by a factor of 4 $\pm$ 1 \citep{atreya2016}. This \{C/H\} can be achieved if the whole planetary core is eroded, but only when the core forms at $r>r_{\rm CO}$, i.e. outside the CO ice line, where a lot of carbon can be stored in the core. Several of our models fit the observed C/H ratio of Jupiter as shown in Fig.~\ref{fig:COplane}. In our simulations, the Jupiters that form in the disk with a lifetime of 5 Myr in the outer regions of the disk have a slightly super solar C/O ratio after core erosion, and a \{C/H\} ratio comparable to observations, independent of the chemical model. However, the Jupiters formed in the disk with a lifetime of 3 Myr do not match \{C/H\} for Jupiter in our solar system, except for one simulation. But this simulation and the ones in the disk that have a lifetime of 5 Myr have a similar starting location: outside the CO ice line ($r_0 > r_{\rm CO}$). From the simulations that match \{C/H\} of Jupiter's atmosphere, we find that \{O/H\} should be increased in similar ways as \{C/H\}, meaning that C/O is roughly solar. This prediction can be tested by the ongoing Juno mission \citep{bolton2006,matousek2007}. On the other hand, disk photoevaporation could also lead to the same predictions, i.e. of similar \{O/H\} enhancement at \{C/H\} and a solar C/O ratio. The degeneracy can be broken by constraints on the presence of a core in Jupiter, which will also be pursued by Juno. While the presence of a significant core and a solar C/O ratio would favor photoevaporation, the lack of a significant core would support core formation by pebble accretion followed by core erosion. 

\section{Discussion} 
\label{sec:discussion} 

In our simulations, the low metallicities of the planetary atmospheres are a direct consequence of the pebble accretion process, which hinders planets to accrete volatiles in solid form during their gas accretion phase. In order to achieve super-solar metallicity of planets, a late stage of planetesimal accretion can be invoked, which we do not model here. On the other hand, as a giant planet migrates slowly (type-II migration) through the disk, it shepherds planetesimals in front of it \citep{tanaka1999} and thus the accretion of planetesimals is not enough to explain a super-solar abundance in carbon or oxygen. The majority of giant planets should thus have a sub-solar composition in carbon and oxygen, except when the planetary core is eroded efficiently and mixed into the atmosphere.

The composition of the accreted material strongly depends on the underlying temperature structure and evolution of the protoplanetary disk (Fig.~\ref{fig:icelines} and Fig.~\ref{fig:growth_tracks}). Here we used the disk model presented in \citet{bitsch2015a}. This model calculates the temperature as a balance between viscous and stellar heating with radiative cooling. The viscous heating is strongly dependent on the viscosity in the disk ($Q^+ \propto \Sigma_{\rm g} \nu$), but viscous heating is only dominant in the inner parts of the disk in the early stages of the disk evolution (see Fig.~\ref{fig:icelines}). This allows the disk to cool down quite quickly and the water ice line crosses the Earth's orbit after 1 Myr, which was also found in \citet{Oka2011} and \citet{Baillie2015}. This rapid cool down poses a problem for the formation history of Earth, as Earth is actually quite dry. Recently \citet{Morbidelli2016} suggested a solution to this problem, by fossilizing the water ice line at its original position due to the growth of Jupiter. As Jupiter's core grows, it blocks the flux of pebbles to the inner disk \citep{lambrechts2014b} and thus hinders ice particles from the outer disk to reach Earth even though the temperature at 1 AU is low enough to allow their existence.

Hot Jupiters in our model form in the inner part of the disk and at rather early times compared to the disk's lifetime \citep{bitsch2015b}, so that the choice of viscosity influences the chemical composition of forming planets. Nevertheless, the initial seeds of most of our hot Jupiters form after $\sim$1 Myr and at a distance of at least 4 AU\footnote{Here $r>r_{\rm H_2 O}$, so water ice is available as pebbles, enhancing the formation of the planetary core}, so the effect on their chemical composition should be minimal. In the outer parts of the disk, starting at a few AU, stellar heating is the dominant heating source, making our temperature model independent of viscosity in the regions where the cold Jupiters form. Therefore our calculations for cold Jupiters and Jupiter in our own solar system should not be affected by the underlying disk model.

Volatiles that condense out in the disk form grains and through mutual sticking pebbles, where the growth time-scale from dust to pebbles increases with orbital distance. As soon as pebbles are formed, they drift inward through the disk, which leads to a local depletion of the dust-to-gas ratio at the location where the pebbles formed \citep[see Fig.1 in ][]{lambrechts2014}. As they drift inwards, they can be accreted by a planetary core, which will therefore inherit the chemical composition of the pebbles that it accreted. However, most of the pebbles drift past the planet into the inner disk where they can cross various ice lines, where they eventually evaporate and release their volatiles into the gas component \citep{ros2013}. This can lead to a local enrichment of volatiles in the gas component that could potentially affect the atmospheric composition of a gas accreting planet to some extent \citep{oberg2016}. This effect would not change the chemical composition of the planetary core, which is governed by the composition of the accreted solid pebbles, because the planetary cores do not migrate significantly across an ice line during core formation. Nevertheless, in order to model this effect accurately, the radial drift of pebbles and their consequent evaporation has to be taken into account. 

A comprehensive formation model would involve detailed chemical evolution of both the pebbles and the gas in the disk simultaneously with the formation and migration of the planet through the disk. Such a model is beyond the scope of the current study where our simulations model the planet growth in detail but the chemistry is incorporated a posteriori based on the output of the simulations, as discussed in section~\ref{sec:model_chem}. The recent simulations by \citet{oberg2016} take the effects of evaporation of pebbles and their subsequent enrichment of the gas disc with volatiles into account, but they do not model the planet growth and migration in detail. Other recent studies have investigated the effect of planetesimals and pebbles ablating in the planetary envelope leading to metal enrichment \citep{pinhas2016,venturini2016} which in turn could accelerate giant planet formation \citep{venturini2016}. A future study thus needs to combine all these effects. 

 \citet{mordasini2016} studied the effects of planetesimal accretion on giant planet composition and spectra using models of planet growth via planetesimals and considering planetesimal ablation in the atmosphere. Their results validate the generally finding across all such studies \citep[e.g.,][]{madhusudhan2014c} that substantial accretion of icy planetesimals in giant planetary envelopes leads to oxygen-rich compositions, as the solid composition is typically dominated by H$_2$O. Those of our present models that consider core erosion, which effectively puts the accreted solids back into the envelope, lead to similarly oxygen-rich compositions. Therefore, observations of metal poor and/or oxygen-poor compositions in giant planetary atmospheres would provide evidence against substantial planetesimal accretion or core erosion. 

\section{Summary}
\label{sec:summary}

We investigate the effect of giant planet formation via pebble accretion and subsequent planetary migration on the atmospheric compositions of giant planets. Our goal is to understand if atmospheric abundances of prominent volatile elements such as oxygen and carbon can be used to constrain the formation and migration mechanisms of giant exoplanets as well as Jupiter in the solar system. To this end we combine the planet growth simulations via pebble accretion and planet migration of \citet{bitsch2015b} with the chemical evolution model of accreting gas giants of \citet{madhusudhan2014c} to model giant planets over a wide range of conditions. As the planet grows, it migrates through the disk and crosses various ice lines due to which the chemical composition of the accreted material also changes with time and location of the planet in the disk, which we account for in our models. In the pebble accretion framework, the accretion of pebbles, and hence solids, is stopped when the planet reaches its pebble isolation mass by opening a partial gap in the disk \citep{lambrechts2014}. When the in-fall of pebbles is stopped, the gas surrounding the forming planet is no longer heated externally and can contract to form a gaseous envelope. Consequently, in the standard pebble accretion paradigm it is the composition of the accreted gas that dominates the envelope and atmospheric composition, unless the core eventually erodes into the envelope over time. 

We study here the influence of the chemical composition of the accreting material on the composition of the planetary atmosphere under different assumptions for the disk chemistry, core erosion, migration paths, and initial conditions. We consider three broad classes of planets, all formed by pebble accretion: (i) hot Jupiters, i.e. Jupiter-mass planets in close-in orbits at $\sim$0.1 AU, (ii) cold Jupiters, i.e. Jupiter analogues, and (iii) super-Jupiters, i.e., with masses of several Jupiter masses. For each class of planets, we run a suite of models with different initial conditions and compute the O and C elemental abundance ratios for two different assumptions for the condensate chemistry in the disk and under assumptions of standard pebble accretion with intact cores and with core erosion. We map the resulting O and C abundances onto the atmospheric C-O plane and identify trends in compositions that can be used to constrain the formation and migration mechanisms of giant exoplanets and Jupiter in the solar system. 

Our results show that atmospheric C/H, O/H, and C/O ratios of hot Jupiters could place important constraints on the  conditions of their formation by pebble accretion and subsequent migration. We find that our model hot Jupiters that  formed via standard pebble accretion, with intact cores, and migrated through the disk result in nearly solar C/H, slightly sub-solar O/H of $\sim$0.7-0.8$\times$ solar, and slightly super-solar C/O of $\sim$0.7-0.8. Hot Jupiters that formed in situ also result in similar compositions. However, hot Jupiters that originally formed as cold Jupiters, i.e. at large orbital separations, via standard pebble accretion but migrated later via disk-free (scattering) mechanisms could have substantially sub-solar C/H and O/H ratios and C/O ratios close to 1. 

On the other hand, for the same initial conditions as above, considering that the cores of the planets can erode over time and mixed into the envelope results in super-solar C/H and O/H ratios and solar or sub-solar C/O ratios, i.e. manifestly oxygen-rich compositions. It is to be noted that other processes, e.g. late stage planetesimal accretion and disk photoevaporation, which we do not model here could also enrich the metallicities. Thus, while observations of sub-solar metallicities and super-solar C/O ratios in hot Jupiters would be strongly indicative of pebble accretion with intact cores and/or disk-free migration, observations of super-solar metallicities and sub-solar C/O ratios could point to a multitude of processes including core erosion, late stage planetesimal accretion, and/or disk photoevaporation. Tentatively, most hot Jupiters show subdued H$_2$O features which could be attributed either to clouds/hazes in the atmospheres and/or low metallicities. In cases where significantly sub-solar ($\lesssim0.7\times$ solar) metallicities and/or super-solar C/O ratios are inferred, our results suggest formation by pebble accretion and migration by disk-free mechanisms.  

Our results also provide important predictions for the compositions of Jupiter in the solar system and extrasolar Jupiter analogues. As alluded to above, we find that the compositions of our Jupiter analogues, which we also refer to as cold Jupiters, are strongly dependent on the assumptions regarding its core. If we assume that the cores formed by pebble accretion are intact then we find significantly sub-solar C/H and O/H ratios and C/O ratios close to 1. On the other hand, if we consider core erosion the C/H and O/H ratios can be significantly super-solar and C/O ratios solar or sub-solar. This result has important implications for the O/H ratio of Jupiter which is currently unknown but is a primary goal of the ongoing Juno mission to Jupiter. Given that the C/H ratio of Jupiter is $\sim$4$\pm$1$\times$ solar \citep{atreya2016}, i.e. significantly super-solar, our current models suggest that if Jupiter formed by pebble accretion it must have undergone substantial core erosion and we predict a similarly enhanced O/H ratio and a nearly solar C/O ratio. However, a similar metal enrichment could also result from alternate processes suggested in the literature such as disk photoevaporation and/or  heavy planetesimal accretion, while keeping the core intact. Therefore, independent constraints on the core mass of Jupiter and its O/H and C/H ratios would be able to break the degeneracies between the different enrichment scenarios. 

Our results provide strong impetus for detailed atmospheric observations of giant exoplanets to determine their atmospheric chemical abundances. As we find in this work, such abundances pave the way to placing powerful chemical constraints on the formation and migration processes of giant exoplanets. Various current and upcoming observational facilities are well poised to make detailed atmospheric abundance measurements for giant exoplanets using high-precision and/or high resolution spectra. Currently, precise molecular abundances, mainly of H$_2$O, have already been reported for a handful of transiting hot Jupiters, using HST WFC3 spectra, and for a few directly-imaged giant planets at large orbital separations using spectra from ground-based facilities. Preliminary constraints on the elemental O/H, C/H, and C/O, ratios from such spectra are suggesting that giant exoplanets likely span a wide range of metallicities and C/O ratios, implying that a diversity of processes might be at play. More observations in the future, and particularly with the JWST and major ground-based spectroscopic platforms, will lead to high-precision constraints on these elemental abundance ratios which in turn could provide deep insights into the formation and migration mechanisms of giant exoplanets.

\section*{Acknowledgments}
B.B.,\,and A.J.\,thank the Knut and Alice Wallenberg Foundation for their financial support. B.B.\, also thanks the Royal Physiographic Society for their financial support. A.J.\,was also supported by the Swedish Research Council (grant 2010-3710), the European Research Council (ERC Starting Grant 278675-PEBBLE2PLANET) and the Swedish Research Council (grant 2014-5775). 

\appendix
\section{Outcomes of planetary growth tracks}
\label{ap:outcomes}

We show in table~\ref{tab:planetdata} several quantities of our planet growth simulations. CJ and HJ indicate if the resulting planet is a hot Jupiter (HJ) or cold Jupiter (CJ), SCJ indicates a cold super-Jupiter and SHJ a hot super-Jupiter,  and IS indicates planets formed in situ. The `Start time' indicates when we put the planetary seed in the protoplanetary disk and the `End time' indicates the end time of the simulation. $r_0$ and $r_{\rm f}$ denote the initial position and final position, respectively, in the disk. $M_{\rm core}$ denotes the core mass of the planet, $M_{\rm env}$ the envelope mass of the planet, and $M_{\rm tot}$ the total mass of the planet. The final column represents the colour code used for this kind of planet in Fig.~\ref{fig:growth_tracks} and Fig.~\ref{fig:COplane}. The tables~\ref{tab:planetchem} and \ref{tab:planetchem_oberg} show the chemical compositions of the planets corresponding to each model in table~\ref{tab:planetdata}. \{C/H\} denotes the carbon to hydrogen ratio compared to solar value of the planetary atmosphere, \{O/H\} the oxygen to hydrogen ratio compared to solar and the columns with subscript `er' denote the values including a $100\%$ core erosion and subsequent mixing into the planetary atmosphere. The C/O denotes the absolute C/O ratio of the envelope; for reference, the solar C/O ratio is 0.55.

{
\begin{table*}
\centering
\begin{tabular}{ccccccccc}
\hline \hline
Type & Start time [Myr] & End time [Myr] & $r_0$ [AU] & $r_{\rm f}$ [AU] & $M_{\rm core}$ [$M_{\rm Jup}$]& $M_{\rm env}$ [$M_{\rm Jup}$]& $M_{\rm tot}$ [$M_{\rm Jup}$]  & color \\ \hline
HJ &       1.92 &       2.98 &      10.80 &       0.10 &      0.020 &       0.98 &       1.00 & red \\
HJ &       1.23 &       2.00 &      12.00 &       0.10 &      0.025 &       0.94 &       0.96 & red \\
HJ &       0.59 &       1.08 &      12.80 &       0.10 &      0.034 &       0.97 &       1.00 & red \\
HJ &       0.65 &       1.13 &       4.00 &       0.10 &      0.019 &       1.05 &       1.07 & red \\
HJ &       1.37 &       2.69 &       4.20 &       0.10 &      0.008 &       0.97 &       0.98 & red \\
\hline
HJ &       0.50 &       1.01 &      10.00 &       0.10 &      0.031 &       1.92 &       1.95 & green \\
HJ &       1.00 &       1.52 &      10.00 &       0.10 &      0.024 &       0.44 &       0.46 & green \\
HJ &       1.50 &       2.24 &      10.00 &       0.10 &      0.020 &       0.70 &       0.72 & green \\
HJ &       2.00 &       2.98 &      10.00 &       0.10 &      0.018 &       0.80 &       0.82 & green \\
\hline
CJ &       1.53 &       3.00 &      21.50 &       2.94 &      0.031 &       1.01 &       1.04 & blue \\
CJ &       1.38 &       3.00 &      25.00 &       4.01 &      0.034 &       1.01 &       1.04 & blue \\
CJ &       1.27 &       3.00 &      27.50 &       4.80 &      0.036 &       1.01 &       1.04 & blue \\
CJ &       1.17 &       3.00 &      29.70 &       5.42 &      0.038 &       1.02 &       1.06 & blue \\
CJ &       1.01 &       3.00 &      33.20 &       6.54 &      0.041 &       1.02 &       1.06 & blue \\
\hline
CJ &       2.48 &       5.00 &      27.00 &       3.11 &      0.029 &       0.99 &       1.01 & magenta \\
CJ &       2.10 &       5.00 &      33.80 &       5.04 &      0.033 &       0.97 &       1.00 & magenta \\
CJ &       1.80 &       5.00 &      39.20 &       6.78 &      0.036 &       0.97 &       1.00 & magenta \\
CJ &       1.26 &       5.00 &      48.30 &       9.67 &      0.041 &       0.97 &       1.01 & magenta \\
\hline
IS &       1.75 &       3.00 &       5.20 &       5.20 &      0.009 &       1.00 &       1.01 & cyan \\
IS &       0.73 &       3.00 &       0.20 &       0.20 &      0.012 &       1.01 &       1.03 & cyan \\
\hline
SJ &       0.60 &       5.00 &      30.00 &       0.51 &      0.045 &       5.60 &       5.65 & black \\
SJ &       0.80 &       5.00 &      43.00 &       2.29 &      0.043 &       2.43 &       2.48 & black \\
SJ &       0.60 &       2.92 &      16.20 &       0.10 &      0.039 &       6.37 &       6.41 & black \\
SJ &       0.20 &       0.90 &       7.20 &       0.10 &      0.041 &       7.69 &       7.73 & black \\
\hline
\end{tabular}
\caption{Planetary growth table}
\label{tab:planetdata}
\end{table*}
}

{
\begin{table*}
\centering
\begin{tabular}{cccccccccc}
\hline \hline
Type & $r_0$ [AU] & $r_{\rm f}$ [AU] & \{O/H\} & \{C/H\} & C/O & \{O/H\}$_{\rm er}$ & \{C/H\}$_{\rm er}$ & C/O$_{\rm er}$ & color \\ \hline
HJ &      10.80 &       0.10 &       0.71 &       0.99 &       0.76 &       2.89 &       1.40 &       0.27 & red \\
HJ &      12.00 &       0.10 &       0.72 &       0.99 &       0.76 &       3.57 &       1.54 &       0.24 & red \\
HJ &      12.80 &       0.10 &       0.75 &       1.00 &       0.73 &       4.49 &       1.71 &       0.21 & red \\
HJ &       4.00 &       0.10 &       0.73 &       1.00 &       0.75 &       2.81 &       0.98 &       0.19 & red \\
HJ &       4.20 &       0.10 &       0.71 &       0.99 &       0.76 &       1.60 &       1.16 &       0.40 & red \\
\hline
HJ &      10.00 &       0.10 &       0.75 &       1.00 &       0.74 &       2.46 &       1.32 &       0.30 & green \\
HJ &      10.00 &       0.10 &       0.77 &       1.00 &       0.71 &       6.46 &       2.08 &       0.18 & green \\
HJ &      10.00 &       0.10 &       0.72 &       0.99 &       0.76 &       3.84 &       1.59 &       0.23 & green \\
HJ &      10.00 &       0.10 &       0.71 &       0.99 &       0.76 &       3.18 &       1.46 &       0.25 & green \\
\hline
CJ &      21.50 &       2.94 &       0.50 &       0.90 &       1.00 &       3.76 &       1.52 &       0.22 & blue \\
CJ &      25.00 &       4.01 &       0.50 &       0.90 &       1.00 &       4.10 &       1.59 &       0.21 & blue \\
CJ &      27.50 &       4.80 &       0.50 &       0.90 &       1.00 &       4.30 &       1.63 &       0.21 & blue \\
CJ &      29.70 &       5.42 &       0.50 &       0.90 &       1.00 &       4.43 &       1.65 &       0.21 & blue \\
CJ &      33.20 &       6.54 &       0.49 &       0.90 &       1.00 &       4.50 &       4.16 &       0.51 & blue \\
\hline
CJ &      27.00 &       3.11 &       0.44 &       0.80 &       1.00 &       3.35 &       3.70 &       0.61 & magenta \\
CJ &      33.80 &       5.04 &       0.39 &       0.72 &       1.00 &       3.77 &       4.09 &       0.60 & magenta \\
CJ &      39.20 &       6.78 &       0.35 &       0.64 &       1.00 &       4.06 &       4.34 &       0.59 & magenta \\
CJ &      48.30 &       9.67 &       0.26 &       0.48 &       1.00 &       4.43 &       4.63 &       0.58 & magenta \\
\hline
IS &       5.20 &       5.20 &       0.50 &       0.90 &       1.00 &       1.48 &       1.09 &       0.41 & cyan \\
IS &       0.20 &       0.20 &       0.80 &       1.00 &       0.69 &       1.78 &       0.99 &       0.31 & cyan \\
\hline
SJ &      30.00 &       0.51 &       0.55 &       0.96 &       0.96 &       1.42 &       1.12 &       0.44 & black \\
SJ &      43.00 &       2.29 &       0.45 &       0.82 &       1.00 &       2.26 &       2.62 &       0.64 & black \\
SJ &      16.20 &       0.10 &       0.73 &       0.99 &       0.75 &       1.39 &       1.11 &       0.44 & black \\
SJ &       7.20 &       0.10 &       0.81 &       1.00 &       0.68 &       1.43 &       1.02 &       0.39 & black \\
\hline
\end{tabular}
\caption{Elemental ratios of planets in table~\ref{tab:planetdata} assuming Case 1 chemistry from table 1.}
\label{tab:planetchem}
\end{table*}
}

{
\begin{table*}
\centering
\begin{tabular}{cccccccccc}
\hline \hline
Type & $r_0$ [AU] & $r_{\rm f}$ [AU] & \{O/H\} & \{C/H\} & C/O & \{O/H\}$_{\rm er}$ & \{C/H\}$_{\rm er}$ & C/O$_{\rm er}$ & color \\ \hline
HJ &      10.80 &       0.10 &       0.72 &       0.87 &       0.67 &       2.81 &       2.01 &       0.39 & red \\
HJ &      12.00 &       0.10 &       0.73 &       0.89 &       0.67 &       3.48 &       2.37 &       0.38 & red \\
HJ &      12.80 &       0.10 &       0.76 &       0.90 &       0.65 &       4.36 &       2.85 &       0.36 & red \\
HJ &       4.00 &       0.10 &       0.76 &       0.90 &       0.66 &       2.49 &       1.62 &       0.36 & red \\
HJ &       4.20 &       0.10 &       0.72 &       0.87 &       0.67 &       1.58 &       1.34 &       0.47 & red \\
\hline
HJ &      10.00 &       0.10 &       0.76 &       0.91 &       0.66 &       2.41 &       1.80 &       0.41 & green \\
HJ &      10.00 &       0.10 &       0.77 &       0.90 &       0.64 &       6.25 &       3.86 &       0.34 & green \\
HJ &      10.00 &       0.10 &       0.73 &       0.88 &       0.67 &       3.74 &       2.51 &       0.37 & green \\
HJ &      10.00 &       0.10 &       0.72 &       0.87 &       0.67 &       3.10 &       2.16 &       0.38 & green \\
\hline
CJ &      21.50 &       2.94 &       0.36 &       0.65 &       1.00 &       3.51 &       2.36 &       0.37 & blue \\
CJ &      25.00 &       4.01 &       0.36 &       0.65 &       1.00 &       3.84 &       2.53 &       0.36 & blue \\
CJ &      27.50 &       4.80 &       0.36 &       0.65 &       1.00 &       4.03 &       2.64 &       0.36 & blue \\
CJ &      29.70 &       5.42 &       0.36 &       0.65 &       1.00 &       4.15 &       2.70 &       0.36 & blue \\
CJ &      33.20 &       6.54 &       0.36 &       0.65 &       1.00 &       4.31 &       4.20 &       0.54 & blue \\
\hline
CJ &      27.00 &       3.11 &       0.32 &       0.58 &       1.00 &       3.21 &       3.46 &       0.59 & magenta \\
CJ &      33.80 &       5.04 &       0.28 &       0.52 &       1.00 &       3.65 &       3.87 &       0.58 & magenta \\
CJ &      39.20 &       6.78 &       0.26 &       0.46 &       1.00 &       3.95 &       4.15 &       0.58 & magenta \\
CJ &      48.30 &       9.67 &       0.19 &       0.35 &       1.00 &       4.33 &       4.48 &       0.57 & magenta \\
\hline
IS &       5.20 &       5.20 &       0.36 &       0.65 &       1.00 &       1.31 &       1.16 &       0.49 & cyan \\
IS &       0.20 &       0.20 &       0.80 &       1.00 &       0.69 &       1.78 &       0.99 &       0.31 & cyan \\
\hline
SJ &      30.00 &       0.51 &       0.53 &       0.74 &       0.76 &       1.37 &       1.19 &       0.48 & black \\
SJ &      43.00 &       2.29 &       0.33 &       0.59 &       1.00 &       2.12 &       2.38 &       0.62 & black \\
SJ &      16.20 &       0.10 &       0.72 &       0.88 &       0.67 &       1.36 &       1.22 &       0.49 & black \\
SJ &       7.20 &       0.10 &       0.82 &       0.94 &       0.63 &       1.36 &       1.18 &       0.48 & black \\
\hline
\end{tabular}
\caption{Elemental ratios of planets in table~\ref{tab:planetdata} assuming Case 2 chemistry from table 1.}
\label{tab:planetchem_oberg}
\end{table*}
}

\bibliographystyle{mnras}
\bibliography{refs_pebbles.bib}

\bsp	
\label{lastpage}
\end{document}